# From Safety-Driven Redundancy to Value-Driven Architecture: An ODD-Based Rationalization Framework for Level 4 Autonomous Vehicles

**Tiago Simão Ferreira**[1,*], **Farzad Fallahi**[2]

*1 Federal Institute of Minas Gerais (IFMG), Santa Luzia, Minas Gerais, Brazil*

*2 University of Luxembourg, Luxembourg*

* Corresponding author: tiago.simao@ifmg.edu.br

## Abstract

Level 4 autonomous vehicles require fault-tolerant architectures capable of preserving critical driving functions under sensing, computing, communication, power, and actuation failures. As these systems progress toward industrial scalability, architecture complexity, validation burden, integration effort, and lifecycle cost become increasingly relevant. This study proposes an Operational Design Domain (ODD)-based, safety-constrained systems-engineering framework for architecture rationalization. ODD-specific capabilities, physical fault assumptions, post-fault operational policies, and spatial sensing requirements are propagated through an integrated safety-feasibility gate. Architecture transformations - including removal, relocation, functional reassignment, and consolidation - are evaluated only within the resulting safety-feasible design space, followed by lifecycle-cost sensitivity and robust multiobjective Pareto analysis. A generic, synthetically parameterized Level 4 architecture is used as a demonstrative computational case study. Within this modeled design space, direct ablation showed limited rationalization potential, whereas directional redistribution and functional consolidation generated a broader set of safety-feasible design candidates. Under the illustrative lifecycle-cost model, the best nominal candidate achieved a 5.55% normalized cost reduction and remained cost-reducing across all 5,000 sampled points in the prescribed economic uncertainty domain. The analysis identified 15 non-dominated architecture-policy solutions representing seven physical transformations. These results support capability-based, policy-conditioned, and lifecycle-aware architecture rationalization under the stated assumptions, while safety remains a non-compensatory feasibility constraint. The reported transformations are design candidates within the modeled assumptions rather than validated production architectures.

## 1. Introduction

Vehicle automation has evolved from isolated driver-assistance functions toward integrated Automated Driving Systems (ADS) capable of assuming perception, decision-making, and vehicle-control responsibilities. In conventional Advanced Driver-Assistance Systems, the human driver generally remains the ultimate fallback. At SAE Level 4, however, the ADS is expected to perform the Dynamic Driving Task (DDT) and manage fallback within its intended Operational Design Domain (ODD) without relying on an immediate human response as the primary safety mechanism [SAE J3016]. This transition changes the role of vehicle architecture: sensing, computing, communication, power supply, and actuation must not only enable nominal operation, but also support fault detection, degraded operation, and achievement of a Minimum Risk Condition.

Redundancy has consequently become a central principle in highly automated vehicles. Cameras, LiDARs, radars, computing nodes, communication paths, power supplies, and actuation channels may be duplicated or diversified to preserve critical capabilities after faults. Nevertheless, fail-operational design does not necessarily require complete duplication of every function. Selective, incomplete, and asymmetric redundancy can preserve required capabilities while limiting architecture complexity (Cassel et al., 2020; Frigerio et al., 2021; Drago et al., 2026). Redundancy should therefore not be interpreted as architectural waste: its value depends on the ODD conditions, fault assumptions, residual capabilities, and fallback functions it supports. This distinction becomes increasingly important as Level 4 systems progress from capability demonstration toward industrial scalability, where hardware cost, E/E integration, wiring, packaging, validation, maintenance, and serviceability become additional design constraints.

The resulting engineering problem is broader than reducing sensor or component count. A component that appears redundant during nominal operation may provide indispensable post-fault capability, whereas an apparently inexpensive duplicated function may generate substantial integration and lifecycle burden. Conversely, functionality may sometimes be relocated, reassigned, or consolidated rather than physically duplicated. This reasoning is consistent with automotive Value Engineering, Design-to-Cost, and DFMA principles, which associate cost with required function rather than with component presence alone. In a safety-critical Level 4 architecture, however, such rationalization must remain subordinate to safety: a lower-cost solution is not an improvement if it violates nominal or post-fault capability requirements. Moreover, the hardware required after a failure depends on the residual mission assigned to the ADS. Preserving degraded DDT capability may require a different physical architecture from executing only a constrained Minimum Risk Maneuver. Post-fault operational policy must therefore be treated as an architecture design variable rather than only as a runtime decision.

Existing research provides many of the individual elements required for this reasoning, but largely in separate methodological streams. ODD specification and monitoring have been linked to functional degradation; spatial ODD characteristics have been translated into sensing range and field-of-view requirements; MRM and graceful-degradation studies quantify post-fault operational behavior; perception research addresses complementary and incomplete sensor redundancy; and fault-tolerant or reconfigurable architectures preserve functions after failures (Jiang et al., 2024; Ye and Wang, 2022; Lee et al., 2024; Aniculăesei et al., 2025). In parallel, automotive E/E research addresses architecture topology, ECU consolidation, zonal organization, wiring reduction, and cost optimization. However, ODD-, perception-, and fallback-oriented approaches generally stop before lifecycle architecture rationalization, while architecture-cost approaches generally do not derive candidate designs from an explicit chain of ODD-specific capabilities, physical faults, residual post-fault requirements, and spatial sensing constraints.

To the authors' knowledge, an integrated architecture-level method has not yet been reported that propagates ODD-specific capability requirements through physical fault assumptions, post-fault residual-capability policies, and spatial sensing constraints; evaluates transformations such as removal, relocation, reassignment, and consolidation; and subsequently identifies robust lifecycle-value alternatives within the resulting safety-feasible design space. Existing methods can determine whether a capability is required, whether a predefined fault-tolerant architecture remains operational, or whether an E/E topology reduces cost, but they do not provide a common decision process for determining whether a specific architecture element can be removed, relocated, or consolidated under explicit ODD-, fault-, post-fault-policy-, and spatial-safety constraints. The gap is therefore not the absence of methods for redundancy, ODD definition, fallback, architecture optimization, or Value Engineering individually, but the lack of an integrated decision process connecting them. The central principle adopted here is consequently non-compensatory: safety defines the feasible design space, whereas Value Engineering searches within that space.

The objective of this study is to develop an ODD-based, safety-constrained systems-engineering framework for the rationalization of Level 4 autonomous-vehicle architectures while preserving specified nominal and post-fault capabilities. The proposed method (i) derives capability requirements from ODD conditions, fault assumptions, and post-fault operational policies; (ii) combines global functional and spatial sensing requirements through an integrated safety-feasibility gate; (iii) evaluates architecture rationalization through removal, relocation, functional reassignment, and consolidation rather than ablation alone; and (iv) applies lifecycle-cost sensitivity and robust multiobjective Pareto analysis only to safety-feasible architecture-policy alternatives. A generic, synthetically parameterized Level 4 architecture is used as a demonstrative computational case study to evaluate the framework's internal behavior, traceability, and decision logic without making claims regarding the architecture, safety case, or production economics of any specific commercial platform.

## 2. Related Work

### 2.1 ODD, Functional Degradation, and Minimum-Risk Operation

The Operational Design Domain (ODD) defines the conditions under which an Automated Driving System is intended to perform the Dynamic Driving Task (DDT). Recent research has progressively extended the ODD from a static deployment descriptor toward an engineering input for runtime monitoring and fallback decisions. ODD-monitoring approaches can identify approaching operational limits and trigger functional degradation before nominal performance becomes unsafe. Jiang et al. (2024), for example, connect ODD monitoring to degraded vehicle behavior, illustrating how operating-domain constraints can be translated into residual functional requirements.

A related research stream addresses DDT fallback, graceful degradation, and Minimum Risk Maneuvers (MRMs). Lee et al. (2024) show that MRM timing and stopping strategy affect post-fault safety outcomes, while Aniculăesei et al. (2025) demonstrate controlled degradation and transfer of decision/control responsibilities when automated operation reaches its limits. Together, these studies establish that post-fault behavior should not be represented simply as an operational/non-operational state. Instead, different faults may require different residual capabilities depending on whether the ADS is expected to continue a degraded DDT or only reach a Minimum Risk Condition.

However, the physical architecture required to satisfy these alternative residual missions is generally not derived simultaneously across sensing, compute, network, power, and actuation. Consequently, fallback policy is well established as an operational-safety problem, but its direct use as a vehicle-architecture design variable remains comparatively underdeveloped.

### 2.2 Perception Safety, Spatial Requirements, and Sensor Redundancy

Multi-sensor perception is a major source of both safety capability and architectural complexity in automated vehicles. Cameras, LiDARs, and radars provide overlapping but non-equivalent information, with different strengths and failure mechanisms across illumination, weather, geometry, and object characteristics. Cassel et al. (2020) formalize this distinction through incomplete multisensor redundancy, in which different sensors preserve different subsets of the required perception information rather than duplicating the complete perception function. Thus, redundancy is more appropriately evaluated by the capabilities and failure conditions it covers than by sensor count alone.

ODD-dependent sensing requirements may also be spatial. Ye and Wang (2022), for example, derive required detection ranges and angles from freeway-entrance geometry and vehicle kinematics, showing that an ODD can impose not only a functional requirement but also a directional sensing requirement. This distinction is relevant because a perception capability may remain globally available while becoming insufficient in a specific region around the vehicle. Global capability availability and spatial coverage therefore represent related but distinct architecture constraints.

Reconfiguration and analytical-redundancy studies further show that required information need not always remain tied to the original physical sensor. Viana et al. (2022) demonstrate degraded localization through alternative sensing strategies following failures, while cross-modal methods such as LiGenCam illustrate how information associated with one sensing modality may, under restricted conditions, be reconstructed from another. These approaches support functional reassignment as a potential architecture mechanism, although they do not establish that physical sensing elements can automatically be eliminated from a production L4 system.

Overall, perception research provides methods for capability and spatial requirements, as well as partial functional reassignment after failures, but does not generally propagate these requirements through the complete vehicle architecture and subsequent lifecycle-value assessment.

### 2.3 Fault-Tolerant Architecture, Reconfiguration, and E/E Rationalization

Fault-tolerant architecture research addresses continuity of critical functions across sensing, computing, communication, power, and actuation. Dependability studies compare redundant and asymmetric hardware structures under explicit fault assumptions, while computer-aided architecture methods evaluate candidate topologies before physical implementation is fixed. Drago et al. (2026) and Julitz et al. (2023), for example, demonstrate increasingly quantitative approaches to fault-tolerant architecture assessment and synthesis. These methods move redundancy design beyond simple hardware duplication, but typically evaluate predefined architecture alternatives rather than deriving them from a complete ODD-fault-residual-policy chain.

Reconfiguration provides an additional mechanism for maintaining functionality after failure. Instead of permanently duplicating all resources, functions may be transferred among sensing, computational, or control resources when a fault occurs. This is consistent with selective and asymmetric redundancy: an architectural element is valuable because of the capability it preserves under a credible fault, rather than because it duplicates another physical element.

In parallel, automotive E/E development has moved from highly distributed ECU architectures toward domain, centralized, and zonal configurations. This transition is strongly associated with ECU consolidation, harness reduction, functional integration, packaging simplification, and centralized computation. Frigerio et al. (2021) compare architecture topologies using cost, failure probability, communication load, and cable length, while Maier and Reuss (2023) address zonal architecture and harness optimization. Broader automotive Value Engineering and DFMA literature similarly emphasizes part-count reduction, functional integration, wiring simplification, and early architecture-level cost decisions.

These studies establish that architecture transformation and simplification are mature engineering concepts. However, consolidation itself is not the objective; it is acceptable only when the required nominal and post-fault capabilities remain satisfied.

### 2.4 Lifecycle Value and the Remaining Integration Gap

Value Engineering, Design-to-Cost, DFMA, and lifecycle-cost methods share a common principle: engineering value should be associated with required function, rather than with preservation of individual components. Component elimination, functional integration, wiring reduction, and architecture simplification are therefore valuable only when required performance, quality, and manufacturability remain preserved. For a Level 4 automated vehicle, the same reasoning must additionally include safety-critical nominal and post-fault functions.

Lifecycle value also extends beyond bill-of-material cost. Integration, validation, calibration, wiring, packaging, maintenance, software support, and transformation-specific engineering effort may materially alter the economic attractiveness of an architecture alternative. A component reduction can therefore remain technically feasible while failing to produce lifecycle value once the cost of implementing and validating the transformation is considered.

The literature reviewed above provides strong methods for individual portions of the architecture-design problem, but these portions remain largely disconnected. ODD and fallback research links operating conditions to residual behavior; perception research links scenarios to sensing capability and spatial coverage; dependability studies evaluate fault-tolerant architecture; and automotive E/E and Value Engineering methods address architecture simplification and lifecycle cost. What remains insufficiently integrated is the complete sequence linking ODD conditions, fault assumptions, post-fault policy, residual capability requirements, spatial sensing constraints, architecture transformation, and lifecycle-value assessment.

Accordingly, the contribution pursued in this study is not a new ODD definition, perception algorithm, fallback strategy, or conventional cost-optimization technique. Instead, the proposed framework acts as an architecture-level integration layer in which these previously separated requirements are propagated through a common decision process. Safety remains a hard, non-compensatory feasibility condition, and Value Engineering is performed only within the resulting policy-conditioned feasible architecture set.

**Table 1. Condensed literature-gap matrix.**

| Research stream / representative work | ODD/C | Fault | Degraded/MRM | Spatial | Arch. transf. | Lifecycle | Robust opt. |
|---|---|---|---|---|---|---|---|
| Cassel et al. (2020) - multisensor safety | ◐ | ✓ | — | ✓ | ◐ | — | — |
| Frigerio et al. (2021) - E/E topology | — | ✓ | ◐ | — | ✓ | ✓ | ◐ |
| Ye & Wang (2022) - ODD sensing geometry | ✓ | — | — | ✓ | — | — | — |
| Viana et al. (2022) - reconfiguration | — | ✓ | ✓ | — | ✓ | — | — |
| Jiang et al. (2024) - ODD degradation | ✓ | — | ✓ | ◐ | — | — | — |
| Lee et al. (2024) - MRM evaluation | ◐ | ✓ | ✓ | — | — | — | — |
| Aniculăesei et al. (2025) - graceful degradation | ◐ | ✓ | ✓ | — | ◐ | — | — |
| Drago et al. (2026) - fault-tolerant architecture | — | ✓ | ◐ | — | ◐ | — | — |
| Automotive VE / E/E optimization literature | — | ◐ | — | — | ✓ | ✓ | ◐ |
| Proposed framework | ✓ | ✓ | ✓ | ✓ | ✓ | ✓ | ✓ |

Legend: ✓ direct coverage; ◐ partial coverage; — not explicitly addressed.

## 3. Methodology

### 3.1 Framework and Problem Formulation

The proposed methodology formulates Level 4 architecture rationalization as a safety-constrained Value Engineering problem. ODD-specific requirements, physical faults, post-fault policy, residual capability, architecture transformation, and lifecycle value are integrated in the sequence summarized in Figure 1. Safety requirements are evaluated before economic objectives; consequently, lifecycle cost cannot compensate for insufficient nominal or post-fault capability.

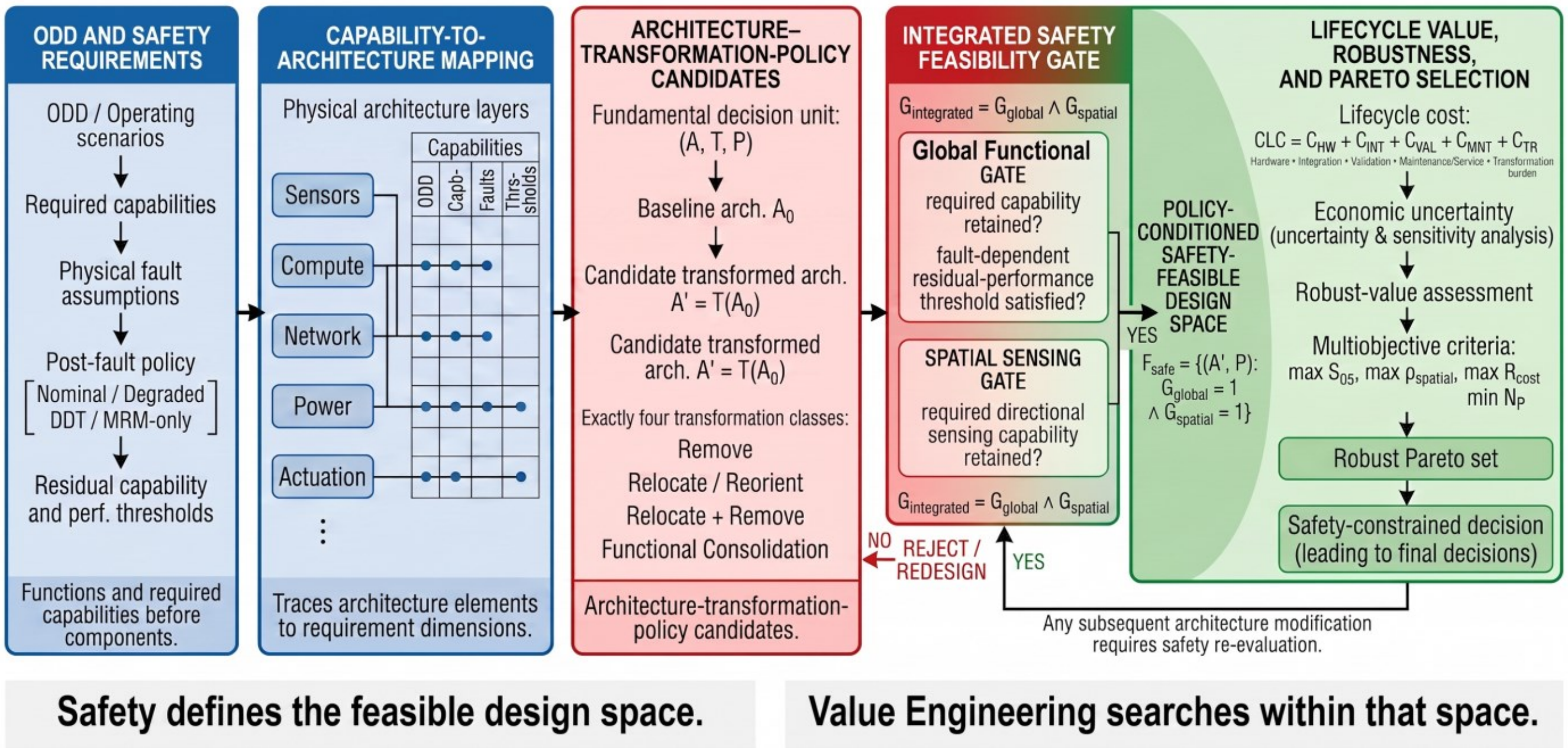

*Figure 1. ODD-based safety-constrained architecture rationalization framework for Level 4 autonomous vehicles.*

Let A denote the physical architecture, T an architecture transformation, and P the post-fault operational policy. The fundamental decision unit is therefore

$$(A, T, P) \tag{1}$$

rather than architecture A alone. This distinction is required because the same physical fault can impose different hardware requirements depending on whether the ADS must preserve degraded DDT capability or only execute a Minimum Risk Maneuver (MRM).

$$\mathcal{F}_{safe} = \left\{(A, P) \colon G_{(global)(A,P)} = 1 \wedge G_{(spatial)(A,P)} = 1\right\} \tag{2}$$

Architecture transformations and economic analysis are performed only within this set. The formulation is therefore explicitly non-compensatory: safety determines admissibility, while lifecycle value ranks only admissible alternatives.

### 3.2 ODD-, Fault-, and Policy-Conditioned Requirements

Nominal functional requirements are represented by the ODD-to-capability matrix

$$R_{o,c}^{OC} = \begin{cases} 1 & \text{if capability c is required in ODD condition o,} \\ 0 & \text{otherwise.} \end{cases} \tag{3}$$

Post-fault requirements are conditioned by an operational policy P(f) ∈ {nominal, degraded DDT, MRM-only}, yielding

$$R_{o,c,f}^{OCF} = \Psi\left(R_{o,c}^{OC}, P(f)\right) \tag{4}$$

where Ψ(·) maps each fault-policy pair to the capabilities that must remain available. Equation (4) explicitly separates fault occurrence from required post-fault mission. Hence, a fault does not determine redundancy by itself; redundancy follows from the residual capability required after that fault.

$$\Theta_{o,c,f} \in [0,1] \tag{5}$$

Each active requirement receives a normalized minimum retained-performance threshold. In an industrial implementation, Θ should be derived from system requirements, simulation, validation, or the vehicle safety case. The demonstrative case uses normalized illustrative values and does not interpret them as certification thresholds. Physical fault categories may contain multiple realizations f^(v)∈F_phys, with architecture-element availability A_f,e∈[0,1]; complete failure corresponds to A_f,e=0, while partial degradation may use intermediate values.

### 3.3 Integrated Global and Spatial Safety Gate

Let E={e₁,...,e_NE} represent the architecture elements and x_e∈{0,1} their presence in candidate architecture A. The normalized contribution of element e to capability c is B_e,c, while M_o,e represents ODD-dependent performance variation.

$$q_{o,c,f,e} = x_e B_{e,c} M_{o,e} A_{f,e} \tag{6}$$

For sensing functions, overlapping contributions are combined through the bounded saturating operator

$$\varphi^{raw}_{o,c,f,sensing} = 1 - \prod_{(e \in E_s)} 1 - q_{o,c,f,e} \tag{7}$$

Equation (7) is used strictly as a bounded engineering aggregation rule representing diminishing marginal contribution; it is not a probability model and does not assume statistical independence among sensors. Correlated degradation and common-cause effects must be modeled separately through the fault and performance models or by replacing this aggregation rule in higher-fidelity implementations.

$$\varphi^{raw}_{o,c,f,l} = \max_{e \in E_l} q_{o,c,f,e}\,, l \neq sensing \tag{8}$$

Layer performance is normalized against the reference architecture $A_0$, and retained system-level capability is determined by the weakest required layer:

$$\Phi_{(o,c,f)(A)} = \min_{l \in L_c} \hat{\varphi}_{(o,c,f,l)(A)} \tag{9}$$

$$\Phi_{(o,c,f)(A)} \geq \Theta_{o,c,f}, \qquad \forall (o,c,f){:}\, R^{OCF}_{o,c,f} = 1 \tag{10}$$

Because functional availability does not guarantee directional sensing availability, a second Safety Gate is introduced. Let C_e,s denote sensor coverage in spatial region s, and $R^{Ocs}$(o,c,s) the corresponding ODD-dependent spatial requirement.

$$q^{sp}_{o,c,f,e,s} = x_e B_{e,c} M_{o,e} A_{f,e} C_{e,s} \tag{11}$$

$$\hat{\varphi}^{sp}_{o,c,f,s} \geq \Theta_{o,c,f} R^{OCS}_{o,c,s} \tag{12}$$

$$G_{integrated} = G_{global} \wedge G_{spatial} \tag{13}$$

$$\rho_{integrated} = \min(\rho_{global}, \rho_{spatial}) \tag{14}$$

This two-level formulation allows a candidate to be rejected even when the required function remains globally available but is insufficient in a safety-relevant spatial region.

### 3.4 Architecture Rationalization Transformations

Only architectures evaluated through the Safety Gate are considered for rationalization. Instead of limiting Value Engineering to component ablation, four transformation mechanisms are evaluated.

$$A^{-e} = A \setminus \{e\} \tag{15}$$

Removal eliminates element e and is accepted only if the transformed architecture-policy pair remains in F_safe.

$$C'_{e,s} = \mathcal{R}_{\theta(C_{e,s})} \tag{16}$$

Relocation or reorientation redistributes, rather than creates, sensing coverage.

$$T(A) = Remove[Relocate(A, e_r, \theta), e_d] \tag{17}$$

A coupled relocation-and-removal transformation tests whether redistributing one element can enable removal of another.

Functional consolidation transfers part of a donor element d to a host h. With transfer coefficient η∈[0,1],

$$B'_{h,c} = 1 - (1 - B_{h,c})(1 - \eta B_{d,c}) \tag{18}$$

$$C'_{h,s} = 1 - (1 - C_{h,s})(1 - \eta C_{d,s}) \tag{19}$$

The donor is subsequently removed, and the transferred contribution inherits the host's ODD modifier and fault availability. Hence, consolidation does not retain artificial independent redundancy from hardware that no longer exists. The coefficient η is an abstract functional-transfer parameter used to explore the design space. In an industrial implementation, it should be calibrated using host-donor capability evidence from sensor characterization, signal-processing performance, geometric coverage, packaging constraints, simulation, and validation. More generally, the scalar may become capability- and scenario-dependent, e.g., η_d,h,c,o,s, when sufficient project data are available.

The transformation search therefore asks whether the required function can be removed, relocated, reassigned, or consolidated while preserving safety feasibility.

### 3.5 Lifecycle Value, Economic Robustness, and Pareto Selection

Economic evaluation is performed exclusively for candidates in F_safe. Normalized lifecycle cost is defined as

$$C_{LC} = C_{HW} + C_{INT} + C_{VAL} + C_{MNT} + C_{TR} \tag{20}$$

where the terms represent hardware, integration, validation, maintenance/service, and transformation-specific engineering burden. The reference architecture is normalized as C_LC($A_0$)=100, and candidate lifecycle saving is

$$\Delta C_i = C_{(LC)(A_0)} - C_{(LC)(A_i)} \tag{21}$$

Thus, reducing hardware does not necessarily imply positive value because integration, host upgrade, software, and validation burden may offset hardware savings.

$$\Delta C_{i,k} = C_{(LC,k)(A_0)} - C_{(LC,k)(A_i)} \tag{22}$$

$$R_{cost,i} = \frac{|\{k: \Delta C_{i,k} > 0\}|}{N_{samples}} \tag{23}$$

R_cost represents robustness only within the explored assumption domain. It is a descriptor of the prescribed uncertainty set and is not interpreted as an empirical probability of commercial cost reduction.

$$R_{cost} \geq 0.95 \wedge S_{05} > 0 \tag{24}$$

$$\Delta C_{min} > 0 \tag{25}$$

Final candidate selection remains multiobjective. For solution i=(A,T,P), the optimization seeks

$$\max S_{05,i}, \max \rho_{spatial,i}, \max R_{cost,i}, \min N_{P,i} \tag{26}$$

$$\mathbb{P}_{RV} = \{(A,T,P) \in \mathcal{F}_{safe}: \nexists j \in \mathcal{F}_{safe}, j \succ i\} \tag{27}$$

The framework therefore produces non-dominated architecture-policy design candidates within the stated requirements and uncertainty domain rather than a single universal optimum or a production-validated architecture.

### 3.6 Computational Implementation

The framework was implemented as a deterministic Python workflow using structured matrices for ODD requirements, capabilities, physical fault realizations, architecture contributions, spatial coverage, and lifecycle assumptions. Candidate architectures are generated through explicit transformation operators and evaluated sequentially through the global and spatial Safety Gates. Economic sensitivity is deliberately decoupled from safety evaluation: once an architecture-policy pair is classified as safety-feasible, changes in economic assumptions cannot alter its safety classification.

The computational model and results were frozen before manuscript preparation to maintain traceability between the formal methodology, demonstrative case study, and reported results.

## 4. Demonstrative Level 4 Case Study

### 4.1 Case-Study Scope, Reference Architecture, and ODD

A generic, synthetically parameterized Level 4 architecture was constructed to demonstrate the framework without relying on proprietary information from any commercial ADS. The case study is a methodological proof of concept rather than a physical validation or production-cost prediction. The reference architecture $A_0$ contains 18 elements distributed across sensing (8), compute (2), network (2), power (2), and actuation (4) layers; its representative layout and the transformation classes evaluated in the study are summarized in Figure 2.

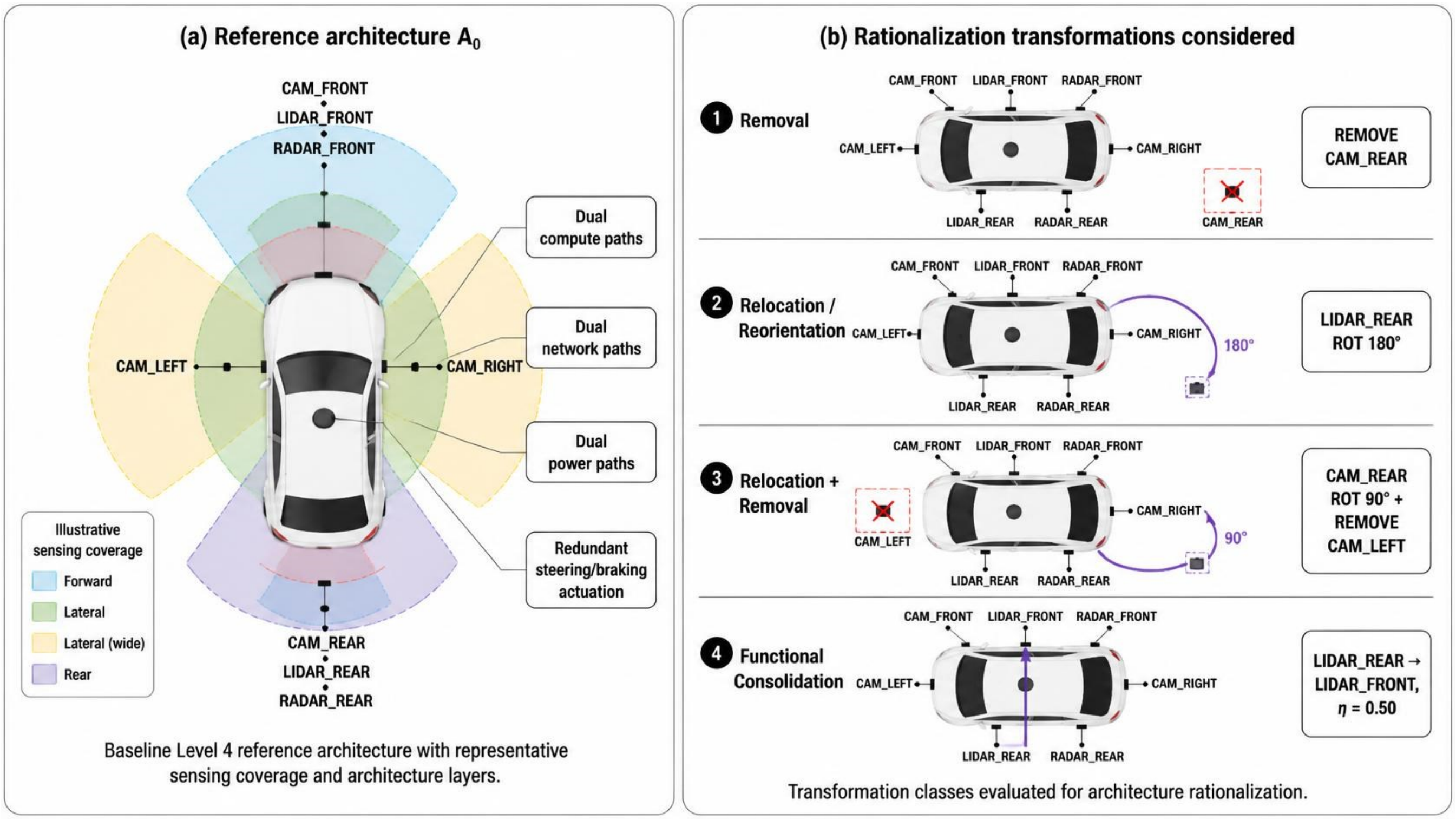

*Figure 2. Reference Level 4 architecture and rationalization transformations considered in the case study.*

$$\mathcal{O} = \{\text{urban_day}, \text{urban_night}, \text{urban_rain}, \text{urban_fog}, \text{urban_intersection}, \text{urban_pedestrian_zone}\} \quad (28)$$

These are evaluation slices rather than mutually exclusive ODD dimensions; real operating conditions may combine several of them simultaneously. Sixteen capabilities were modeled across perception, localization, planning, control, and safety/fallback functions. The five road-based ODD slices require 13 nominal capabilities, whereas the pedestrian-zone slice requires 12 because lane-boundary detection is not imposed. These requirements constitute an illustrative systems-engineering specification rather than a universal Level 4 capability set.

### 4.2 Faults, Post-Fault Policies, and Residual Requirements

$$F = \{\text{nominal, camera, lidar, radar, misalignment, contamination,} \\ \text{compute, network, latency, power, steering, braking}\} \quad (29)$$

**Table 2. Baseline fault-response policy.**

| Response | Fault categories |
|---|---|
| Nominal | Nominal operation |
| Degraded DDT | Camera, LiDAR, radar, misalignment, contamination, compute-node, network-path, and power-path failures |

| Response | Fault categories |
|---|---|
| MRM-only | Excessive network latency, steering-actuator failure, braking-actuator failure |

The degraded-DDT mode retains the complete road-driving capability set plus fault-management functions, while MRM-only operation requires a reduced subset sufficient to execute a controlled transition toward a Minimum Risk Condition. To evaluate the effect of operational policy on architecture requirements, the five sensing-related fault categories {camera, lidar, radar, misalignment, contamination} were subsequently allowed to switch between degraded-DDT and MRM-only requirements.

Fault categories were expanded into 18 physical realizations to distinguish failures of individual redundant compute, network, power, steering, and braking paths. Complete failures use A_f,e=0, while the illustrative partial-degradation factors were

$$A_{misalignment} = 0.70, \qquad A_{contamination} = 0.60, \qquad A_{latency} = 0.60 \tag{30}$$

**Table 3. Normalized residual-performance requirements.**

| Capability group | Nominal | Degraded DDT | MRM-only |
|---|---|---|---|
| Perception | 0.85 | 0.70 | 0.50 |
| Localization | 0.90 | 0.80 | 0.60 |
| Planning | 1.00 | 0.80 | 0.60 |
| Control | 1.00 | 0.90 | 0.80 |
| Safety / fallback | 1.00 | 1.00 | 1.00 |

These values express retained performance relative to $A_0$; they are not absolute safety probabilities or certification thresholds.

### 4.3 Functional and Spatial Sensing Parameterization

Camera, LiDAR, and radar were assigned normalized capability profiles reflecting their complementary roles. For reproducibility, the eight sensing-capability profiles used in the case study were

**Table 4. Normalized sensing capability profiles.**

| Capability | Camera | LiDAR | Radar |
|---|---|---|---|
| Object detection | .75 | .90 | .75 |
| Object classification | .90 | .55 | .30 |
| Range estimation | .45 | 1.00 | .85 |
| Relative velocity | .40 | .55 | 1.00 |
| Free space | .75 | .95 | .35 |
| Lane boundary | .90 | .75 | 0 |
| Ego localization | .50 | .70 | .30 |
| Ego motion | .60 | .60 | .60 |

**Table 5. ODD-dependent sensing modifiers.**

| ODD condition | Camera | LiDAR | Radar |
|---|---|---|---|
| Day | 1.00 | 1.00 | 1.00 |
| Night | .70 | 1.00 | 1.00 |
| Rain | .60 | .75 | .95 |
| Fog | .45 | .60 | .90 |
| Intersection | 1.00 | 1.00 | 1.00 |
| Pedestrian zone | 1.00 | 1.00 | 1.00 |

These coefficients are normalized engineering assumptions intended to create modality-dependent degradation; they are not empirical sensor-performance curves.

$$\mathcal{S} = \{front, rear, left, right\} \tag{31}$$

The directional coverage coefficients C_e,s used in this four-sector abstraction are summarized in Figure 3(b); Figure 3(a) shows the corresponding sensor-to-sector interpretation.

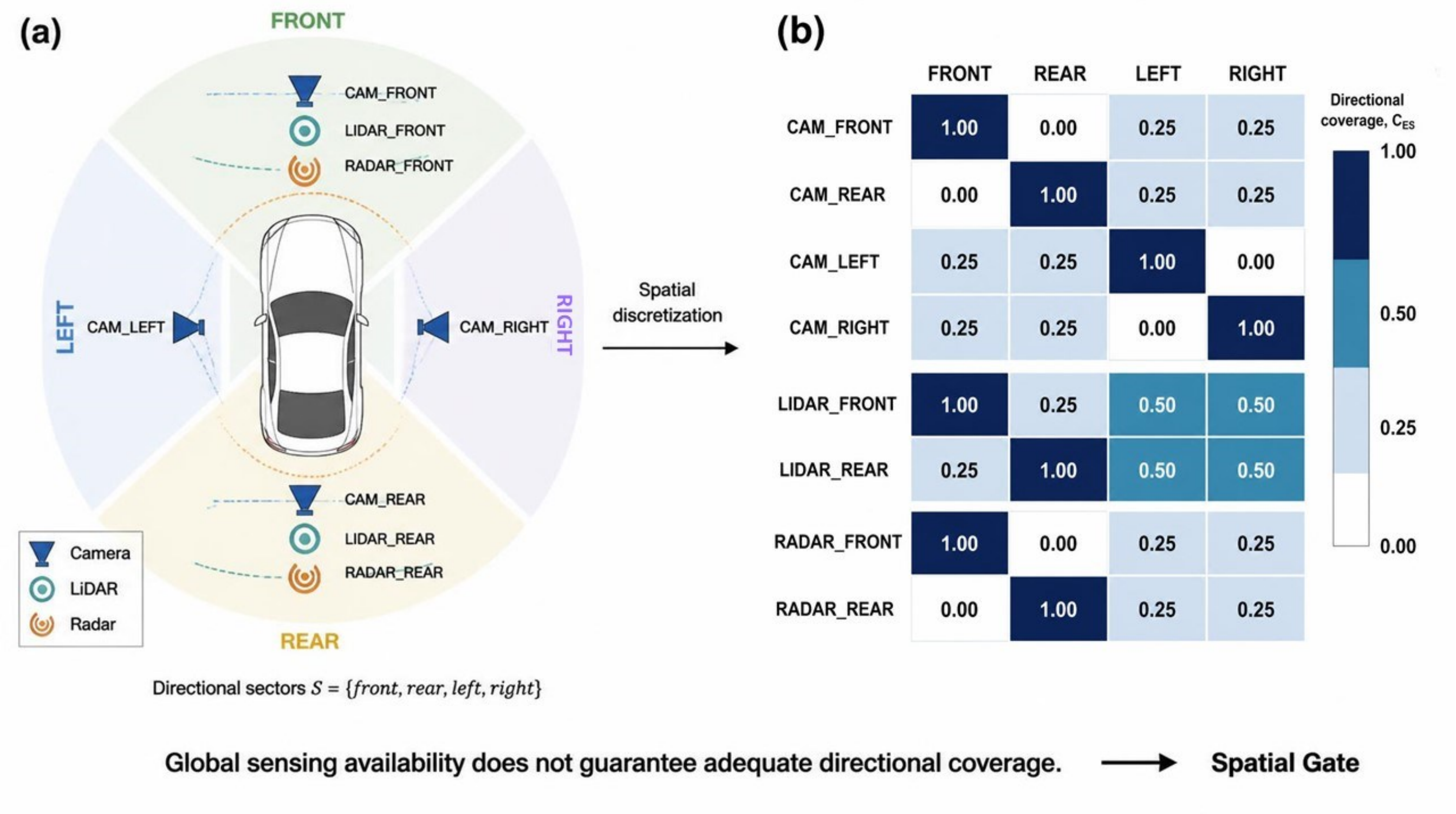

*Figure 3. Spatial sensing abstraction and directional coverage coefficients adopted in the demonstrative case study.*

Only the six environment-relative perception capabilities were spatially resolved. The four-sector abstraction is intentionally coarse and serves to distinguish global functional availability from directional availability, not to reproduce production sensor geometry. Industrial use would replace C_e,s with scenario- or geometry-based visibility, range, field-of-view, and detection-performance models.

### 4.4 Transformation Space and Lifecycle-Cost Uncertainty

The case study evaluates the four transformation mechanisms defined in Section 3. Direct removal was applied to all 18 elements, followed by pairwise removal search. For sensing redistribution, each of the eight sensors was evaluated at $\theta \in \{90°, 180°, 270°\}$, giving 24 relocation candidates before policy variation. Feasible relocations were subsequently coupled with removal of a second element. Same-modality consolidation was evaluated for camera-camera, LiDAR-LiDAR, and radar-radar donor-host pairs at

$$\eta \in \{0.50, 0.75, 1.00\} \tag{32}$$

yielding 48 donor-host-transfer candidates before policy variation. Cross-modality consolidation was excluded because its quantitative representation would require additional empirical assumptions. The scalar η is used only as a demonstrative design-space parameter and does not establish physical equivalence between donor and host sensors.

$$C_{(LC)(A_0)} = 100 \tag{33}$$

Illustrative normalized hardware weights were Camera=1, LiDAR=6, Radar=2.5, HPC=8, Network=2, Power=1.5, Steering=4, and Braking=4. Integration, validation, maintenance, and transformation-specific burdens were then applied by architecture layer; therefore, component elimination does not automatically recover its full hardware contribution.

Economic robustness was evaluated over a 12-dimensional uncertainty space covering sensor hardware, sensing integration/validation/maintenance, removal and relocation burden, and consolidation host-upgrade, integration, and validation assumptions. Multiplier ranges varied broadly around nominal values, typically from approximately 0.5-0.6 to 1.4-2.0.

$$N_{samples} = 5000 \tag{34}$$

These ranges are not statistical distributions derived from production data. Accordingly, the robustness metrics quantify stability within the prescribed assumption domain, not the probability of achieving a particular production cost reduction. The economic model also does not monetize the operational-availability consequences of more restrictive post-fault policies, such as mission interruption, recovery, towing, or fleet downtime. Post-fault policy burden is therefore retained as a separate Pareto objective rather than converted into lifecycle cost.

### 4.5 Interpretation Boundaries

The numerical case study is designed to test the behavior, internal consistency, traceability, and decision logic of the proposed framework within stated assumptions, not to determine the optimal architecture of an existing Level 4 platform. The ODD slices, capability requirements, thresholds, sensing coefficients, environmental modifiers, four-sector geometry, fault-severity parameters, functional-transfer coefficients, lifecycle weights, and uncertainty bounds are intentionally synthetic and normalized.

A feasible transformation should therefore be interpreted as a rationalization design candidate under the stated assumptions.

Consequently, a transformation classified as feasible should not be interpreted as evidence that an equivalent physical component can be removed from a commercial ADS. Likewise, lifecycle savings demonstrate the behavior of the Value Engineering and robustness layers within the modeled economic domain; they are not supplier-price forecasts, empirical commercial confidence intervals, or probabilities of production cost reduction. For industrial application, the demonstrative inputs would need to be replaced by project-specific ODD and safety requirements, measured or simulated perception performance, physical fault and dependability models, detailed geometric sensing requirements, host-donor integration evidence, lifecycle-cost data, and operational-availability costs associated with degraded operation or MRM-only policies. The bounded sensing aggregation, four-sector spatial representation, and scalar functional-transfer coefficient are replaceable model instances within the framework rather than universal Level 4 design models.

## 5. Results

### 5.1 Baseline Feasibility and Policy-Conditioned Safe Set

The reference architecture $A_0$ satisfied all global functional requirements under the baseline policy but failed the spatial Safety Gate in six cases, with a minimum spatial margin of -0.1337. The largest deficit occurred for front-sector relative-velocity estimation under radar-family failure in urban fog, demonstrating that global functional availability alone can mask directional insufficiency.

The five sensing-related fault categories were then varied between degraded-DDT and MRM-only requirements, producing 32 policy combinations. Integrated feasibility of $A_0$ was first recovered when camera-, LiDAR-, and radar-family failures were assigned to MRM-only operation; four policies in total made $A_0$ feasible, and two of those also admitted removal of CAM_REAR. Figure 4 summarizes this policy-conditioned safe set. Under the baseline policy, individual sensor removals produced 292 incremental spatial failures; assigning all five sensing-related fault categories to MRM-only reduced this count to 73, approximately 75% fewer.

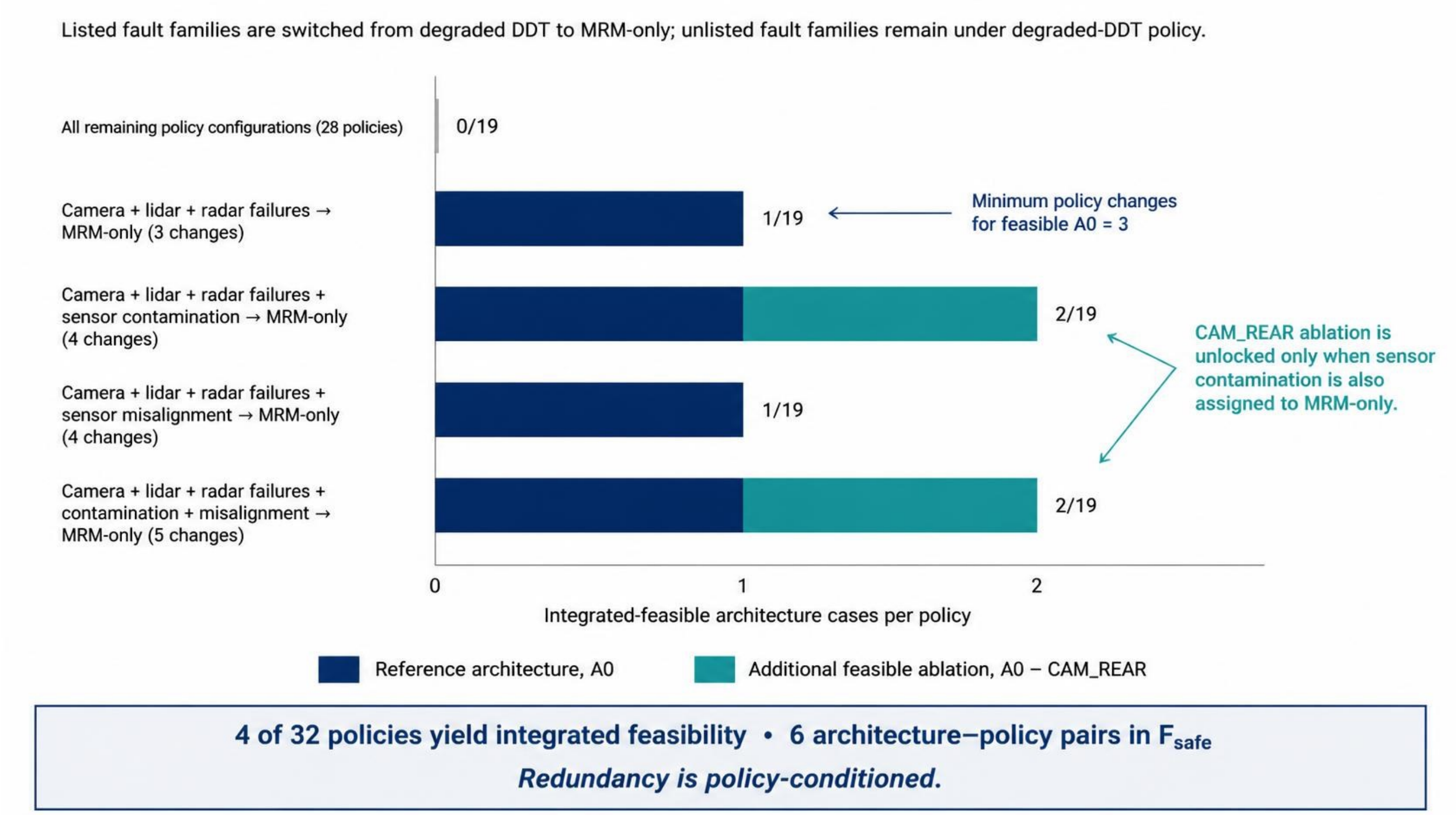

*Figure 4. Effect of post-fault operational policy on integrated safety feasibility.*

### 5.2 Architecture Rationalization: From Ablation to Functional Transformation

Direct ablation yielded almost no rationalization opportunity: CAM_REAR was the only individually removable element, and no feasible two-element removal was found across the 612 pairwise cases evaluated under the four $A_0$-feasible policies. Relocation changed the result: 40 of 96 relocation-policy combinations remained integrated-feasible, and a 180° reorientation of LIDAR_REAR increased the spatial margin from 0.0126 to 0.0863.

Redistribution also altered the downstream removal space. Of 680 coupled relocation-removal cases, 24 were integrated-feasible and 18 unlocked removals that direct ablation could not achieve. Same-modality functional consolidation produced the largest feasible set: 130 of 192 candidate-policy cases were integrated-feasible and 112 enabled donor removal that had been infeasible under direct ablation. Figure 5 therefore shows the central progression from ablation-only reasoning toward spatial and functional redistribution.

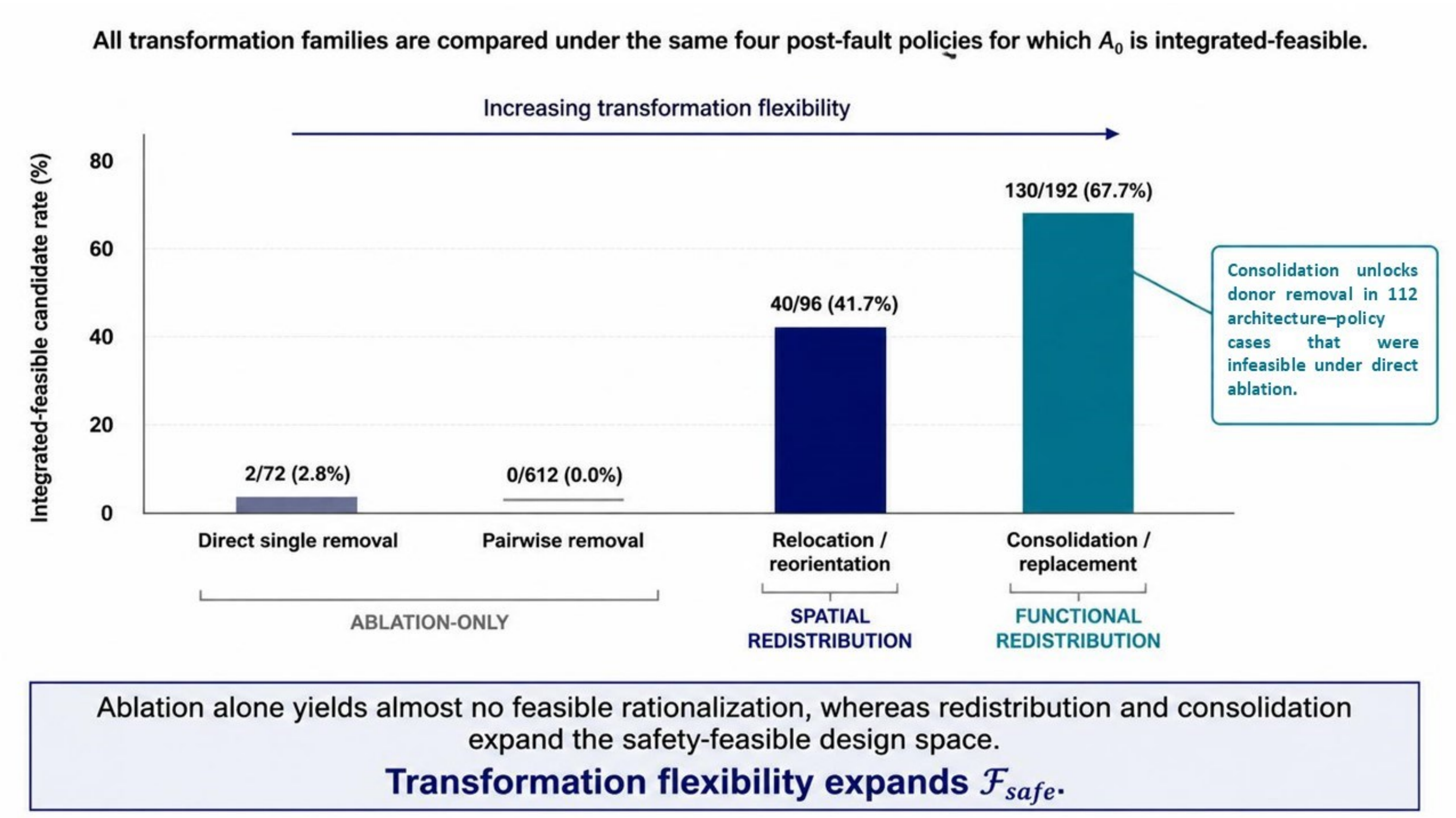

*Figure 5. Expansion of the safety-feasible design space with increasing transformation flexibility.*

These counts depend on the demonstrative parameterization and do not establish physical sensor substitutability; rather, they show that an ablation-only search can substantially underestimate the architecture-transformation space.

### 5.3 Lifecycle Value and Economic Robustness

Only integrated safety-feasible candidates were submitted to lifecycle-value analysis. With $C_LC(A_0)=100$, the best nominal candidate was partial consolidation of LIDAR_REAR into LIDAR_FRONT at $\eta=0.50$, yielding a normalized lifecycle saving of 5.5491 and a spatial margin of 0.0848 under a four-change post-fault policy.

Economic robustness was then evaluated over 5,000 Latin Hypercube samples of the 12-dimensional uncertainty domain. As illustrated in Figure 6, representative removal, relocation-removal, and consolidation candidates remained cost-reducing throughout the sampled assumption space but differed substantially in saving magnitude and dispersion. The LiDAR consolidation candidate achieved S05=2.2464, median saving 4.7759, and minimum sampled saving 0.7489.

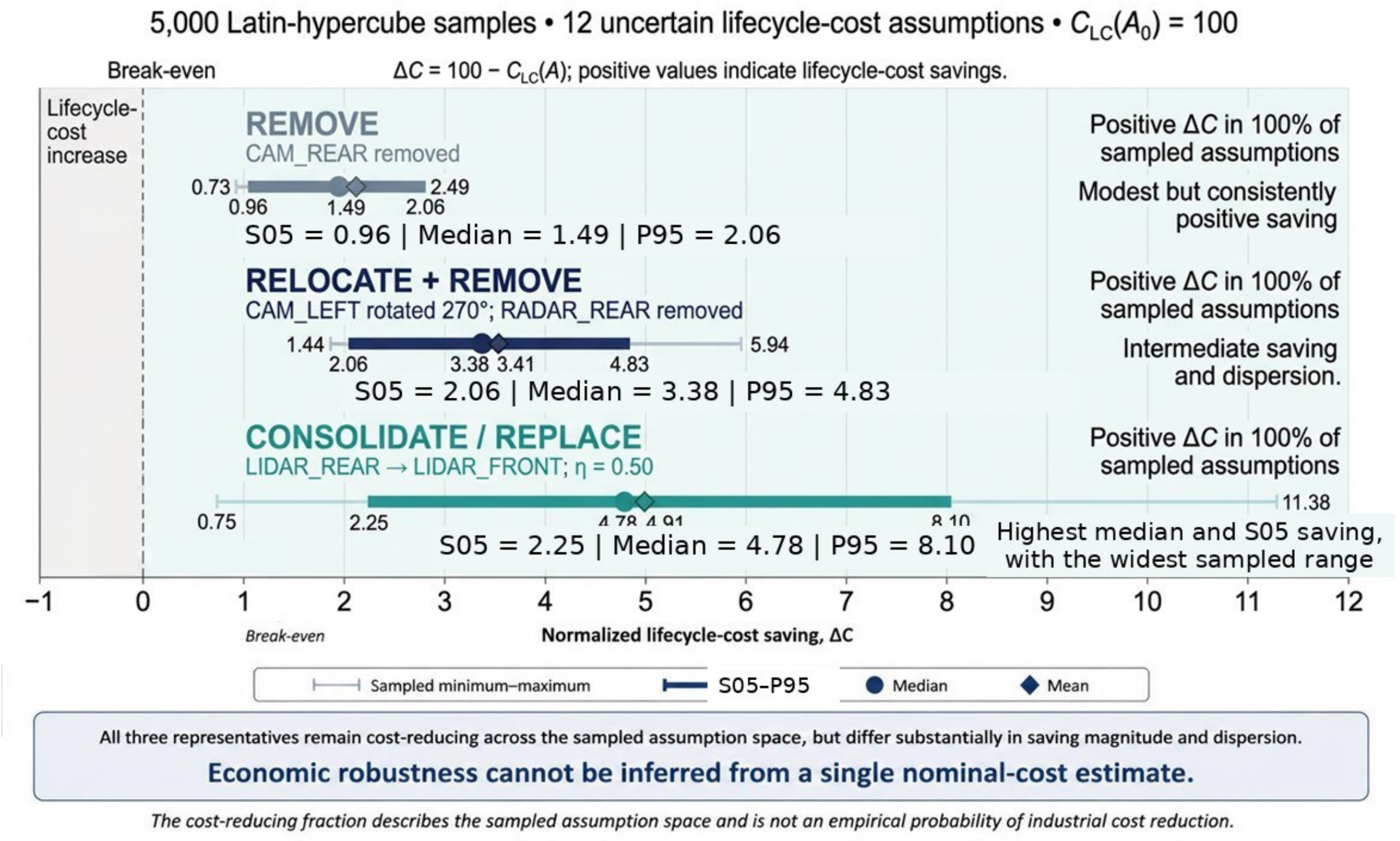

*Figure 6. Economic robustness of representative safety-feasible architecture transformations under lifecycle-cost uncertainty.*

For this candidate, the dominant economic influences were LiDAR hardware cost (Spearman $\rho$=+0.789), consolidation validation burden ($\rho$=-0.400), and host-upgrade burden ($\rho$=-0.268). Accordingly, robustness here denotes stability within the prescribed assumption domain and does not represent an empirical probability, commercial confidence level, or guarantee of production cost reduction.

### 5.4 Robust Architecture-Policy Pareto Set

The final optimization returned to the complete decision unit (A,T,P). Of 156 integrated safety-feasible transformed architecture-policy solutions, 150 met the high-frequency robust-value criterion and 50 remained cost-reducing across the complete sampled economic domain. The four-objective search identified 15 non-dominated architecture-policy solutions representing seven physical transformations.

Figure 7 shows that no single candidate simultaneously maximized robust lifecycle saving, residual spatial margin, and policy simplicity. Partial LIDAR consolidation at $\eta$=0.50 provided the largest fifth-percentile saving (S05=2.2464), whereas partial RADAR_REAR-to-RADAR_FRONT consolidation reached a spatial margin of 0.1500 with fewer policy changes.

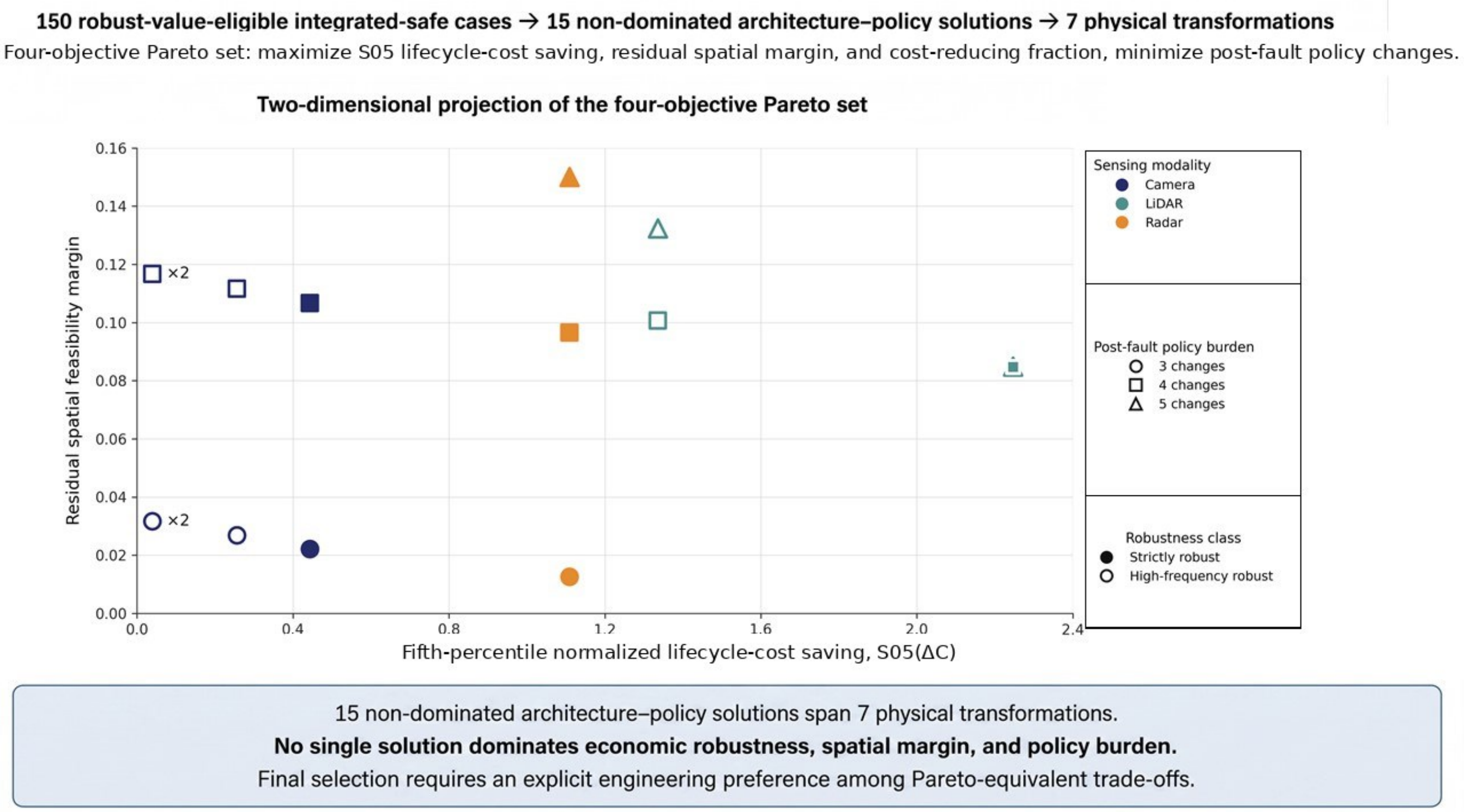

*Figure 7. Robust-value Pareto front of integrated safety-feasible architecture-policy solutions.*

The policy-conditioned nature of the trade space remained visible at the Pareto stage: increasing the admissible policy burden from three to four changes increased the maximum S05 from 1.1091 to 2.2464, while a fifth change did not improve the best robust economic saving but increased the maximum attainable spatial margin. Within the modeled design space, the framework therefore returns a family of non-dominated architecture-policy design candidates rather than a universal minimum-cost architecture.

## 6. Discussion

The results support the study objective within the modeled design space by demonstrating that architecture admissibility depends on the residual mission assigned after a fault. The same physical baseline that failed the integrated gate under the original degraded-DDT policy became feasible when selected sensing failures were reassigned to MRM-only operation (Figure 4). Under the stated assumptions, redundancy is therefore policy-conditioned: its necessity follows from the combination of ODD, fault, required residual capability, and post-fault policy rather than from hardware duplication alone.

The transformation results further demonstrate, within the modeled design space, why component ablation is an incomplete Value Engineering strategy for fail-operational systems. Direct removal provided almost no opportunity, whereas directional redistribution and functional consolidation expanded the feasible design space (Figure 5). The relevant engineering question is consequently whether the required function can be removed, relocated, reassigned, or consolidated while preserving the applicable global and spatial safety requirements. The consolidation results remain architecture-design candidates under the stated assumptions rather than evidence of direct physical substitutability; in industrial use, the transfer coefficient $\eta$ would require calibration from sensor performance, signal processing, geometry, packaging, and validation evidence.

A third result concerns the separation between safety feasibility and economic preference. Hardware reduction did not automatically imply lifecycle value because integration, validation, maintenance, and host-upgrade burdens could offset hardware savings. Figure 6 demonstrates that economic robustness cannot be inferred from a single nominal estimate: safety-feasible candidates differed materially in both saving magnitude and dispersion across the prescribed uncertainty domain. These results characterize robustness within that modeled domain and do not establish an empirical probability of equivalent commercial savings. Safety therefore remains a non-compensatory gate, with economic analysis applied only inside the admissible set.

The Pareto analysis reinforces the decision-support character of the framework. The non-dominated solutions span different combinations of robust saving, residual spatial margin, robustness class, and post-fault policy burden (Figure 7). No universal optimum emerges because architecture and fallback policy are coupled engineering choices. This decision structure could support industrial architecture reviews once populated with project-specific safety, performance, integration, and cost evidence; the present demonstrative results do not establish production-scale optimality.

The study remains a demonstrative computational case using synthetic, normalized inputs. The six ODD slices, capability contributions, residual thresholds, environmental modifiers, partial-fault severities, four-sector spatial representation, lifecycle-cost coefficients, and uncertainty bounds were selected to exercise the method rather than reproduce a production platform. The bounded sensing aggregation is an engineering approximation, and the four-sector model does not reproduce real field-of-view geometry, range, resolution, occlusion, or time-dependent visibility. In addition, the lifecycle model does not monetize the operational-availability consequences of more restrictive post-fault policies, including mission interruption, recovery, towing, or fleet downtime. Accordingly, the specific transformations and savings reported here should not be generalized to commercial Level 4 architectures.

These limitations constrain quantitative transferability and the strength of industrial claims more than the structure of the proposed decision process. Industrial application would require replacing the illustrative matrices with project-specific ODD requirements, scenario databases, measured or simulated perception performance, geometric visibility models, FMEA/FTA/FMEDA-derived fault assumptions, detailed E/E architecture data, calibrated donor-host transfer relationships, enterprise lifecycle-cost information, and operational-availability costs associated with post-fault policy. Within these boundaries, the results support architecture rationalization as a capability-based, policy-conditioned, safety-constrained, and lifecycle-aware decision process: safety defines what may be changed, and Value Engineering ranks which admissible changes are worth further engineering validation.

## 7. Conclusions

The objective of this study was to develop an ODD-based, safety-constrained systems-engineering framework for rationalizing Level 4 autonomous-vehicle architectures while preserving specified nominal and post-fault capabilities. This objective was addressed by integrating ODD requirements, physical faults, post-fault operational policy, global and spatial safety gates, architecture transformations, lifecycle value, uncertainty, and Pareto selection within a single non-compensatory decision process. The resulting decision unit (A,T,P) makes explicit that architecture requirements depend on both physical configuration and the residual mission assigned after a fault.

Within the demonstrative design space, the results support the first two methodological propositions. Global feasibility alone was insufficient to guarantee directional sensing adequacy, and post-fault policy materially changed the integrated feasible set. Direct ablation provided almost no rationalization opportunity, whereas relocation and functional consolidation expanded the admissible transformation space. These findings support a capability-based rather than component-count-based interpretation of redundancy under the stated requirements and parameterization.

The lifecycle and robustness stages support the remaining objective of identifying value-oriented design candidates without relaxing safety requirements. Safety-feasible transformations differed in nominal saving, uncertainty sensitivity, residual spatial margin, and policy burden; consequently, the final Pareto set comprised multiple non-dominated architecture-policy design candidates under the modeled assumptions rather than a universal or production-validated optimum. Economic benefit therefore ranks alternatives only after mandatory safety constraints have been satisfied.

The quantitative results are a methodological proof of concept rather than predictions for a commercial ADS because the sensing, fault, transfer, spatial, and lifecycle parameters were synthetically normalized. Economic robustness applies only to the prescribed uncertainty domain, and the present lifecycle model does not monetize post-fault operational availability or recovery costs. Industrial deployment would require project-specific safety requirements, measured or simulated performance data, detailed fault and geometric models, calibrated functional-transfer relationships, lifecycle-cost information, and explicit operational consequences of fallback policy. Within these boundaries, the demonstrative results support the study objective and the internal decision logic of Level 4 architecture rationalization as a capability-based, policy-conditioned, safety-constrained, and lifecycle-aware engineering process.

### References

1. **SAE International**, *Taxonomy and Definitions for Terms Related to Driving Automation Systems for On-Road Motor Vehicles*, SAE Recommended Practice J3016_202104, revised April 2021. doi:10.4271/J3016_202104.

2. **Cassel, A., Bergenhem, C., Christensen, O.M., Heyn, H.-M., Leadersson-Olsson, S., Majdandzic, M., Sun, P., Thorsén, A., and Trygvesson, J.**, "On Perception Safety Requirements and Multi Sensor Systems for Automated Driving Systems," *SAE International Journal of Advances and Current Practices in Mobility*, 2(6):3035–3043, 2020. doi:10.4271/2020-01-0101.
3. **Frigerio, A., Vermeulen, B., and Goossens, K.G.W.**, "Automotive Architecture Topologies: Analysis for Safety-Critical Autonomous Vehicle Applications," *IEEE Access*, 9:62837–62846, 2021. doi:10.1109/ACCESS.2021.3074813.
4. **Drago, M., Bondavalli, A., Lollini, P., Antlanger, M., and Drenkelforth, S.**, "Model-Based Dependability Assessment of Fault-Tolerant Architectures for Autonomous Driving Systems," *Journal of Systems and Software*, 241:112972, 2026. doi:10.1016/j.jss.2026.112972.
5. **Singh, V.K., Kumar, R., and Arya, M.K.**, "Automotive Product Development Lifecycle Optimization through Value Engineering and Value Analysis (VAVE) Techniques," in *2017 International Conference on Advances in Mechanical, Industrial, Automation and Management Systems (AMIAMS)*, pp. 349–354, 2017. doi:10.1109/AMIAMS.2017.8069236.
6. **Naiju, C.D.**, "DFMA for Product Designers: A Review," *Materials Today: Proceedings*, 46:7473–7478, 2021. doi:10.1016/j.matpr.2021.01.134.
7. **Jiang, Z., Pan, W., Liu, J., Han, Y., Pan, Z., Li, H., and Pan, Y.**, "Enhancing Autonomous Vehicle Safety Based on Operational Design Domain Definition, Monitoring, and Functional Degradation: A Case Study on Lane Keeping System," *IEEE Transactions on Intelligent Vehicles*, 9(10):6552–6566, 2024. doi:10.1109/TIV.2024.3370836.
8. **Ye, X. and Wang, X.**, "Operational Design Domain of Automated Vehicles at Freeway Entrance Terminals," *Accident Analysis & Prevention*, 174:106776, 2022. doi:10.1016/j.aap.2022.106776.
9. **Lee, M., Jeong, K., Kim, B.-S., Lim, T.-H., Yun, K., and Han, D.**, "A Study on Driving Safety Evaluation Methods for Minimal Risk Maneuver of Automated Driving System," *Transactions of the Korean Society of Automotive Engineers*, 32(12):1015–1023, 2024. doi:10.7467/KSAE.2024.32.12.1015.
10. **Aniculaesei, A., Aslam, I., Zhang, M., Buragohain, A., Vorwald, A., and Rausch, A.**, "Improving Safety of Autonomous Vehicles: A Verifiable Method for Graceful Degradation of Decision and Control Responsibilities," *SAE International Journal of Connected and Automated Vehicles*, 8(2):297–315, 2025. doi:10.4271/12-08-02-0021.
11. **Viana, K., Zubizarreta, A., and Diez, M.**, "A Reconfigurable Framework for Vehicle Localization in Urban Areas," *Sensors*, 22(7):2595, 2022. doi:10.3390/s22072595.
12. **Xu, M., Gu, Y., Goncharenko, I., and Kamijo, S.**, "LiGenCam: Reconstruction of Color Camera Images from Multimodal LiDAR Data for Autonomous Driving," *Sensors*, 25(14):4295, 2025. doi:10.3390/s25144295.
13. **Julitz, T.M., Tordeux, A., and Löwer, M.**, "Computer-Aided Design of Fault-Tolerant Hardware Architectures for Autonomous Driving Systems," *Proceedings of the Design Society*, 3:1047–1056, 2023. doi:10.1017/pds.2023.105.
14. **Maier, J. and Reuss, H.-C.**, "Design of Zonal E/E Architectures in Vehicles Using a Coupled Approach of k-Means Clustering and Dijkstra's Algorithm," *Energies*, 16(19):6884, 2023. doi:10.3390/en16196884.
15. **Asiedu, Y. and Gu, P.**, "Product Life Cycle Cost Analysis: State of the Art Review," *International Journal of Production Research*, 36(4):883–908, 1998. doi:10.1080/002075498193444.